\documentclass[12pt,a4paper,twoside]{article}
\usepackage{epsfig}
\usepackage{graphics}
\newlength{\figwidth}
\newlength{\figheight}
\newlength{\figheightE}
\newlength{\figheightEe}
\newlength{\figheightD}
\newlength{\figheightDm}
\def\pyver{6.214}

\def\gaga{ \gamma \gamma }
\def\bbbar{ b\overline{b} }
\def\ccbar{ c\overline{c} }

\def\gagabb{ \gaga \rightarrow \bbbar }

\def\gagabbcc{ \gagabb , \, \ccbar }

\def\gagabbg{ \gagabb (g) }

\def\gagabbgccg{ \gagabbg , \, \ccbar(g) }

\def\Wgaga{ W_{\gaga} }

\def\epem{ e^{+} e^{-} }
\def\emem{ e^{-} e^{-} }

\def\sqrtsee{ \sqrt{s_{ee}} }
\def\sqrtseeeq{$ \sqrtsee  = $ }

\newcommand{\gagahad}{\( \gaga \rightarrow \) {\em hadrons}}

\newcommand{\bbtagging}{\( \bbbar  \)-tagging}
\newcommand{\btagging}{\( b \)-tagging}

\newcommand{\ccmistagging}{\( \ccbar  \)-mistagging}
\def\AO{A}
\def\HO{H}

\def\MAO{ M_{\AO} }
\def\MAOeq{$ \MAO = $ }

\def\AOHO{ \AO,\HO }

\def\AHbb{ \AOHO \rightarrow \bbbar }
\def\Abb{ \AO \rightarrow \bbbar }
\def\Hbb{ \HO \rightarrow \bbbar }

\def\gagaAbb{ \gaga \rightarrow \Abb }
\def\gagaHbb{ \gaga \rightarrow \Hbb }
\def\gagaAHbb{ \gaga \rightarrow \AHbb }

\def\sgagaAHbb{ \sigma (\gagaAHbb)}

\def\MSSMpars{$ \tan \beta = 7 , \, M_{2} = \mu = $ 200 GeV}

\def\ie{{i.e.\ }}
\title{
       \vspace{-1.5cm}
       \begin{flushright}
       \begin{tabular}{l}
       {\large IFT - 20/2003}    \\[-3mm]
       {\large hep-ph/ }\\[3mm]
       {\large July 2003}
       \end{tabular}
       \end{flushright}
\vspace{1.5cm}
Measurement of the MSSM Higgs-bosons production \\
in \( \gamma \gamma \rightarrow A,H \rightarrow b\overline{b} \) \\
at the Photon Collider at TESLA 
}

 \author{ \\[1cm]
  Piotr Nie\.zurawski, Aleksander Filip \.Zarnecki \\
 {\small\it Institute of Experimental Physics, Warsaw University, 
    ul. Ho\.za 69, 00-681 Warsaw, Poland} \\[3mm]
 Maria Krawczyk \\
 {\small\it Institute of Theoretical Physics, Warsaw University, 
        ul. Ho\.za 69, 00-681 Warsaw, Poland} \\[-2mm]  } 

\date{}

\begin{document} 

\maketitle 

\vfill

\begin{abstract}

Results from a realistic simulation of the heavy MSSM Higgs-bosons $A$ and $H$ 
production,
\( \gamma \gamma  \rightarrow A,H \rightarrow b\overline{b} \), 
at the Photon Collider at TESLA are reported. In the scenario where a light 
SM-like Higgs boson $h$ exists, we study (following M.\ M\"uhlleitner et al. \cite{MMuhlleitner})
Higgs bosons $A$ and $H$ for masses \( M_{A}=200,250,300,350 \) GeV and \(\tan\beta=7\).
This scenario corresponds to parameters region not accessible at LHC 
or at the first stage of the $e^+ e^-$ collider.
NLO estimation of background, analysis of overlaying events, 
realistic $b$-tagging and corrections for escaping neutrinos were performed.
The statistical precision of the cross-section measurement is estimated to be 8-20\%.

\end{abstract}

%%%%%%%%%%%%%%%%%%%%%%%%%%%%% NZK part %%%%%%%%%%%%%%%%%%%%%%%%%%%%%%%%
\newpage

%\begin{center} 

%{%
%\Large
%
%}
%
%\vspace{1cm}
%
%\end{center} 

\section{Introduction}

A photon collider option of the TESLA collider \cite{TDR} offers a  unique possibility to produce
the Higgs boson as an \( s \)-channel resonance. 
The neutral Higgs boson couples to the photons  through a loop with   massive charged particles. 
This loop-induced higgs-$\gaga$ coupling
is sensitive to contributions of new particles, which appear in various 
extensions of  the Standard Model (SM). 
Besides precision measurements, a  photon collider is a candidate for the discovery machine \cite{Asner}. 

In case of Minimal Supersymmetric extension of the SM (MSSM) the 
photon collider will be able 
to cover so called ``LHC wedge'' around intermediate values of 
$ \tan \beta$, $ \tan \beta \sim 7 $,
and for heavy neutral Higgs-bosons masses above 200 GeV,
which will be inaccessible at LHC \cite{ATLAS} and at the first stage of $\epem$ linear colliders.
In our analysis, we consider a SM-like scenario where the lightest Higgs
boson $h$ has properties of the SM-Higgs boson, 
while heavy neutral Higgs bosons
are degenerated in mass and have negligible couplings to the gauge bosons
$W/Z$. 
We take   MSSM parameters as in \cite{MMuhlleitner},
 \ie \MSSMpars, and  consider the process \( \gagaAHbb \) at the Photon Collider 
at TESLA for Higgs-boson masses \MAOeq 200, 250, 300 and 350 GeV.
The MSSM Higgs bosons $\AOHO$ with masses in this range are expected to decay 
into \( \bbbar \) state with branching ratios from around 80\% to 20\%. 

The preliminary  results for the MSSM Higgs-bosons were presented in  
\cite{NZKamsterdam}.
In our previous works \cite{NZKhbbm120appb,NZKSMeps2003} we have considered
a light SM Higgs-boson production at the Photon Collider 
at TESLA. The background treatment introduced  in that analysis  
we apply now for the  MSSM Higgs-bosons.
Both, the signal and  background
events are generated according to a  realistic photon--photon luminosity
spectrum \cite{V.Telnov}, parametrized by a CompAZ model \cite{CompAZ}. 
This study takes into account overlaying events \gagahad{} 
for which full simulated photon--photon luminosity spectra \cite{V.Telnov} 
are used.
We simulate the detector response according to the program SIMDET 
\cite{SIMDET401} and perform a realistic \btagging{} \cite{Btagging}.

\section{Photon--photon luminosity spectra}

The Compton back-scattering of a laser light off the  high-energy
electron beams is considered as a  source of the highly  energetic, 
highly polarized photon beams  \cite{Ilya}. 
In photon--photon beams simulations for a photon collider option at TESLA,
 according to the  current design \cite{TDR}, the energy of the laser photons
%, \( \omega _{\textrm{laser}} \), 
is assumed to be fixed for all considered electron-beam energies;
the laser photons are assumed to have circular polarization $P_{c} = $  100\%,
while the electrons longitudinal polarization is  $P_{e} = $ 85\%. 
In considered case the luminosity spectrum is peaked at high energy and we assume that 
the energy of primary electrons is adjusted in order to  enhance the signal
at a particular mass.

In a generation of the processes \gagahad, one has to take into account 
also the low energy events, since they contribute to overlaying events.
 To simulate  them   the  realistic $\gaga$-luminosity spectra
for the photon collider at TESLA \cite{V.Telnov} are used, 
with the non-linear corrections and higher order QED processes.
For generation of the processes $\gagaAHbb$ and $\gagabbgccg$ we use the 
CompAZ parametrization \cite{CompAZ} of the $\gaga$-luminosity spectrum \cite{V.Telnov}.
%   W-gg ?
%

%
The results presented in this paper were obtained 
for an integrated luminosity
expected for one year of the photon collider running for TESLA collider 
\cite{V.Telnov}. For example with \sqrtseeeq 419 GeV 
-- the optimal choice for \MAOeq 300 GeV --
the total photon--photon luminosity per year is $L_{\gamma\gamma}=808$ fb$^{-1}$.
%For the considered range of masses 200 - 350 GeV we use luminosities equal
%to 570 fb$^{-1}$ till 937 fb$^{-1}$, for  corresponding  $\emem$-collision energies
%\sqrtseeeq 305--473 GeV.
%
% FZ
The total photon-photon luminosity increases with energy from about 570 fb$^{-1}$ for 
\sqrtseeeq 305 GeV (\MAOeq 200 GeV) 
%the best choice of beam energy for 
to about 937 fb$^{-1}$ for \sqrtseeeq 473 GeV (\MAOeq 350 GeV).
  
%
%and luminosity for $W_{\gamma\gamma}>80$ GeV  ! for m_h=120 GeV
%is $L_{\gamma\gamma}(W_{\gamma\gamma}>80)=84$ fb$^{-1}$.
%

\section{Details of a simulation \\ and the first results for \MAOeq 300~GeV}

To calculate the total widths and  branching ratios of the MSSM Higgs
bosons $\HO$ and $\AO$ we use the program HDECAY \cite{HDECAY} 
with MSSM parameters as in \cite{MMuhlleitner}, \ie \MSSMpars,
and including decays and loops of supersymmetric particles. 

In a generation of the signal events both $\AO$ and $\HO$
are included, due to their degeneracy in mass.
This was done with
the PYTHIA \pyver{} program \cite{PYTHIA}.
A parton shower algorithm, implemented in PYTHIA,
was used to generate the final-state particles.

The  background events due to processes 
$\gagabbgccg$
were  generated using the program written by G.~Jikia \cite{JikiaAndSoldner},
where a complete  NLO QCD  calculation for the production of  massive
quarks is performed within the massive-quark scheme. 
The program includes exact one-loop QCD corrections to the lowest order
(LO) process $\gagabbcc$ \cite{JikiaAndTkabladze}, and in addition 
the non-Sudakov form factor in the double-logarithmic
approximation, calculated up to four loops \cite{MellesStirlingKhoze}.

The fragmentation into hadrons for all processes was performed 
using the PYTHIA program. 

Because of a large cross section,
about two \gagahad{} events\footnote{%
We consider only photon--photon events with $\Wgaga >4 $ GeV.}
are expected on average per bunch crossing at the TESLA Photon-Collider 
(for \( \sqrt{s_{ee}} \approx \) 400 GeV, at nominal 
luminosity).
We generate these events according to PYTHIA \pyver
% where the VMD, direct and GVMD/anomalous events are included.
%
and convolute with results of a full simulation of the photon--photon 
luminosity spectra \cite{V.Telnov}, rescaled for the chosen beam energy.
For each considered $\emem$ energy, $\sqrtsee$, 
an average number of the \gagahad{} events 
per a bunch crossing is calculated.  
Then, for every signal $\gagaAHbb$ or background $\gagabbgccg$ event, 
the \gagahad{} events are overlaid (added to the event record)
according to the Poisson distribution.  

Program  SIMDET version 4.01 \cite{SIMDET401} 
was used  to simulate a detector performance. 
Because \gagahad{} process  has a forward-peaked distribution 
(see for example \cite{NZKamsterdam})
we decrease the influence of overlaying events
by ignoring tracks and clusters with 
$|\cos(\theta_{i})|>\cos(\theta_{min})=0.9$ 
($\theta_{min}=450$ mrad; 
the polar angle is measured in the laboratory frame). 
This cut is used only when overlaying events are included in the analysis.

The jets were reconstructed
using  the Durham algorithm, with \( y_{cut} = 0.02 \); the distance measure
was defined as 
\( y_{ij}=2\min (E^{2}_{i},E^{2}_{j})(1-\cos \theta _{ij})/E^{2}_{vis} \),
where $E_{vis}$ is defined as the total energy measured in the detector.

The following selection cuts were used  to 
%suppress a background:
select the signal events, $ \gagaAHbb $:
\begin{itemize}
\item since the Higgs bosons are expected to be produced
  almost at rest, we require that the ratio of the total 
 longitudinal momentum of all observed particles 
 to the total visible energy is
\( |P_{z}|/E_{vis}<0.15 \),
\item we select two- and three-jet events, \( N_{jets}=2,\, 3 \), so that 
 events with one additional jet due to a hard-gluon emission are also accepted,
\item for each jet we require 
   \( |\cos \theta _{i}|<0.75 \), $i=1, ..., N_{jets}$.
\end{itemize}

We use  ``ZVTOP-B-Hadron-Tagger'' package for 
the TESLA collider \cite{Btagging} for realistic \btagging{} simulation.
The package is based on the neural network algorithm trained on the $Z$ decays. 
For each jet it returns a ``$b$-tag'' value -- the number 
between 0 and 1 corresponding to ``$b$-jet'' likelihood.
In order to optimize the signal cross-section measurement,  
we choose the two-dimensional cut on $b$-tag values for 2-jet events.
For 3-jet events three possible pairs of jets were considered and 
the event was accepted if at least one pair passed the $b\bar{b}$ cut.
It was found that the cut optimal for $\sgagaAHbb$ measurement,
including effects of overlaying events,  
corresponds to the \bbtagging{} efficiency $\varepsilon_{bb}=79\%$
and  \ccmistagging{} probability $\varepsilon_{cc}=3.6\%$.
However, if  overlaying events are not included then the best choice 
corresponds to the efficiences 
$\varepsilon_{bb}=72\%$ and  $\varepsilon_{cc}=1.7\%$.  
%

%%%
\begin{figure}[thb]
{\centering \resizebox*{!}{\figheightE}%
  {\includegraphics{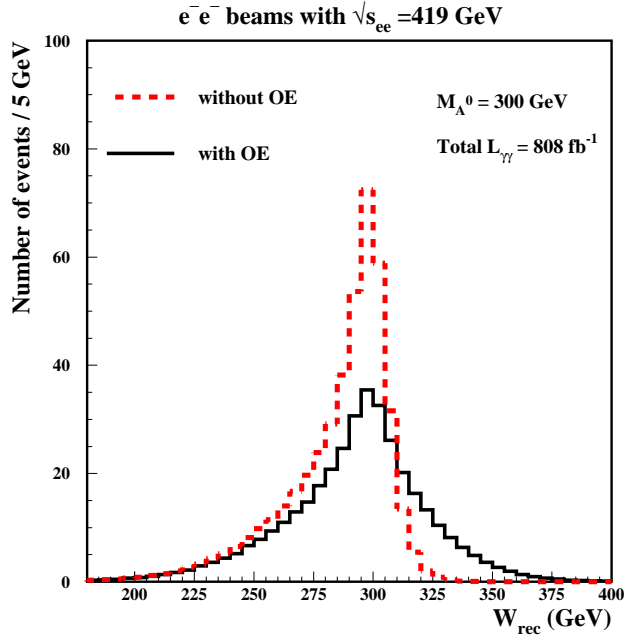}} \par}

\caption{\label{fig:HHOEBtag}
Reconstructed invariant-mass, \protect\( W_{rec}\protect \),
distributions for selected $\gagaAbb$ events  for \MAOeq 300 GeV, after \btagging,
 without and with overlaying events (OE).
}
\end{figure}
%%%

The invariant-mass distributions 
for the $\gagaAbb$  events after \btagging,
before and after taking into account overlaying events (OE),
are compared in Fig.\ \ref{fig:HHOEBtag}  for \MAOeq 300 GeV.
The mass resolutions from the 
Gaussian fits, in the region \( \mu - \sigma  \)
to \( \mu + 2 \sigma  \), to both distributions  are 8 and 20 GeV, respectively. 
The same resolutions are obtained for the selected $\gagaHbb$  events (not shown).
As the expected mass difference between  $\HO$ and $\AO$ is 
of the order of 1~GeV, it will not be possible to 
% distinguish between 
separate signals of scalar and pseudoscalar Higgs boson  using the the invariant-mass distribution.
In the following we will  consider a simultaneous  measurement of the
total cross section for both processes  $\gagaAbb$ and $\gagaHbb$.

%
% for signal-with-overlaying-events than for signal-only events is present.
%
% Here numbers:
%
%
%  NO OE
%  Histo id: 1100111
%  Initial number of events: 905.895
%  Current number of events: 439.25
%  Efficiency              : 0.48488
%
%
%
%  WITH OE
%  Histo id: 11100111
%  Initial number of events: 905.895
%  Current number of events: 367.558
%  Efficiency              : 0.40574
%
%
%
Despite our choice of a $\bbbar$ selection cut 
which corresponds 
to the greater \bbtagging{} efficiency than for the simulation without overlaying events,
we observe a 16\% drop of number of events   (from about 440 to 370 events)
for simulation which includes overlaying events. 
This is because energy deposits from the \gagahad{} processes,
remaining after the $\theta_{min}$ cut, ``shift'' 
jets nearer to the beam axis and the event 
can be rejected by the jet-angle cut.
Moreover, the additional deposits and  $\theta_{min}$-cut deform jets
slightly what reduces a selection efficiency.
%
% To study this issue in details we plan to simulate signal events 
% with various $\theta_{min}$ values. 
%

\begin{figure}[p]
{\centering \resizebox*{!}{\figheightD}%
            {\includegraphics{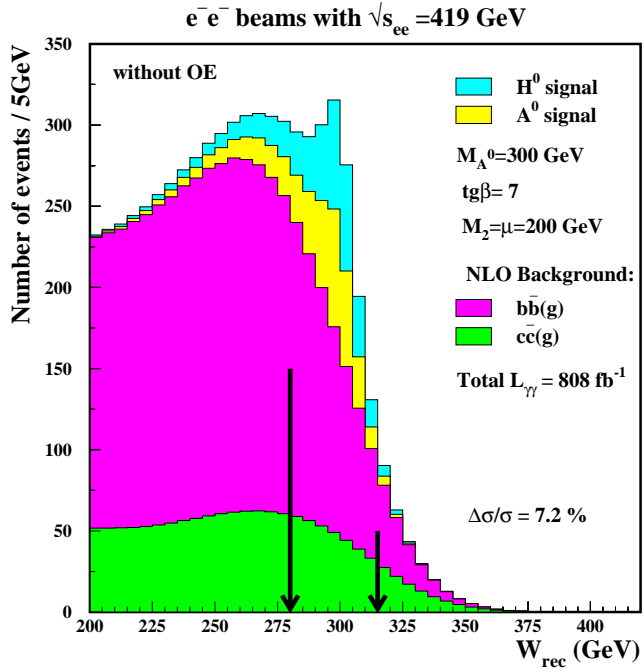}} \par}
{\centering \resizebox*{!}{\figheightD}%
            {\includegraphics{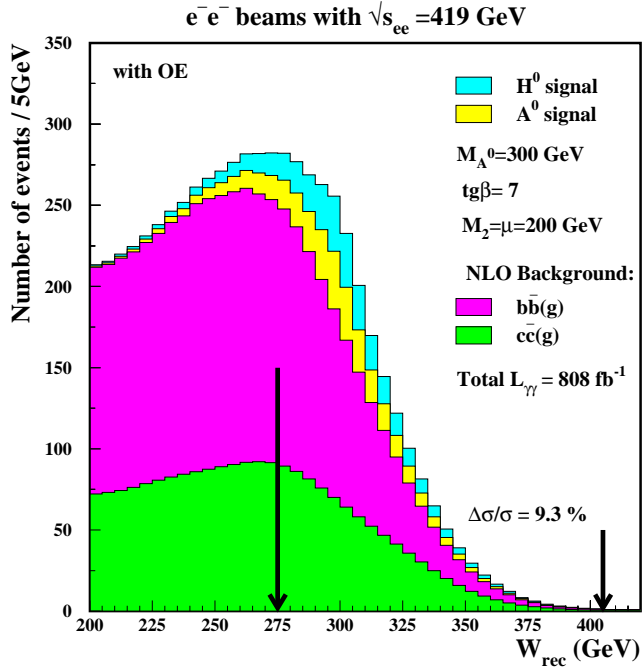}} \par}

\caption{\label{fig:ResultWithNLOBackgd}
Distribution of the reconstructed invariant mass, 
\protect\( W_{rec}\protect \), for the selected $\bbbar$ events  (\MAOeq 300 GeV)
before (upper plot) and after (lower plot) taking into account 
overlaying events (OE). Contributions of the $H$ and $A$ signal 
and of the heavy-quark background are shown separately. 
Arrows indicate the mass windows optimized for the measurement of 
$\sgagaAHbb$.
}
\end{figure}

After applying selection cuts described above and including \btagging{}, 
we obtain the distributions of the reconstructed \( \gaga  \) invariant
mass, \( W_{rec} \), shown in Figs.\ \ref{fig:ResultWithNLOBackgd}.
The $\Hbb$ and $\Abb$ signal, and the NLO background contributions,
$\bbbar(g)$ and  $\ccbar(g)$, are shown separately.
Result obtained before (upper plot) and after (lower plot) taking into account 
overlaying events are compared. 
We observe that the overlaying events significantly smear out the 
Higgs-boson signal.

Assuming that the signal for the  Higgs-bosons production will be extracted
by counting the number of \( \bbbar \) events in the mass window 
around the peak, $N_{obs}$, 
and subtracting the expected background events, $N_{bkgd}$,
we can calculate 
the expected relative statistical error for the cross section 
\( \sgagaAHbb \)
in the following way:
\[
\frac{\Delta \sgagaAHbb}{\sgagaAHbb} =
\frac{\sqrt{N_{obs}}}{N_{obs}-N_{bkgd}}.
\]
For Higgs-boson mass of 300~GeV,
after taking into account overlaying events,
the accuracy of 9.3\% is expected for the reconstructed invariant-mass
window between 275 and 405 GeV 
(see  Fig.\ \ref{fig:ResultWithNLOBackgd}).

\section{Final results for masses 200--350 GeV}

%
%As in \cite{NZKhbbm120appb} to reduce the low invariant-mass tail in
%the reconstructed $\Wgaga$ distribution for the signal events 
%we use the corrected invariant-mass variable, defined as: 
As in \cite{NZKhbbm120appb} to correct for escaping neutrinos
we use the corrected invariant-mass variable, defined as: 
\begin{eqnarray*}
%\label{eq:Wcorr} 
W_{corr} & \equiv &  \sqrt{W^{2}_{rec}+2P_{T}(E_{vis}+P_{T})} \; .
\end{eqnarray*}

\begin{figure}[thb]
{\centering \resizebox*{!}{\figheightEe}%
            {\includegraphics{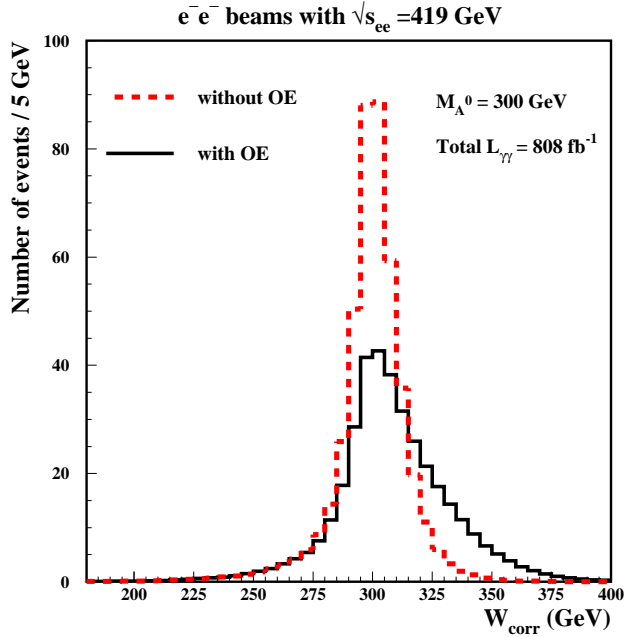}} \par}

\caption{\label{fig:HHOEBtagWcorr}
Corrected invariant mass, \protect\( W_{corr}\protect \),
distributions for the selected $\gagaAbb$ events for \MAOeq 300 GeV, after \btagging, 
obtained without and with overlaying events (OE).
}
\end{figure}

\begin{figure}[bht]
{\centering \resizebox*{!}{\figheightEe}%
            {\includegraphics{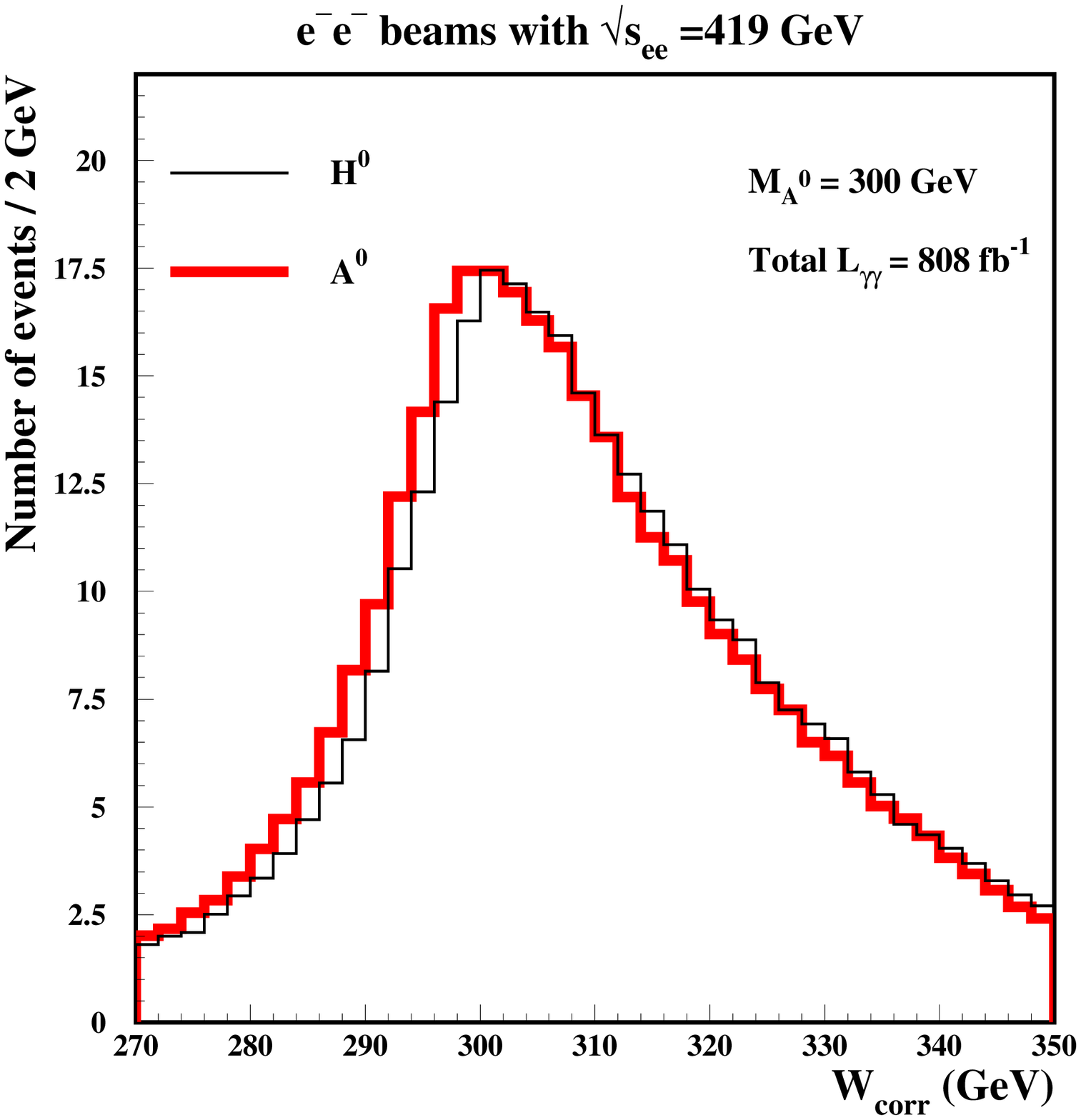}} \par}

\caption{\label{fig:AHBtagWcorr}
Corrected invariant mass, \protect\( W_{corr}\protect \),
distributions for the selected $\gagaHbb$ and $\gagaAbb$ events for \MAOeq 300 GeV, after \btagging, 
with overlaying events.
}
\end{figure}

In Fig.\ \ref{fig:HHOEBtagWcorr} the distributions of \( W_{corr} \)
for the selected $\gagaAbb$ events, obtained without and with overlaying events, 
are presented. 
The tail of events with the invariant masses below $\sim 280$ GeV 
is much smaller than for $W_{rec}$ (compare with Fig.\ \ref{fig:HHOEBtag}). 
The mass resolutions, derived from the 
Gaussian fits to the $W_{corr}$ distributions 
in the region from \( \mu - 2 \sigma  \)
to \( \mu + \sigma  \),
without and with overlaying events,
 are equal to 8 and 13 GeV,  respectively. 
For $\gagaHbb$ events we obtain a very similar distribution, as seen in Fig.\ \ref{fig:AHBtagWcorr}.

\begin{figure}[thb]
% \vspace{-1cm}
{\centering \resizebox*{!}{\figheightDm}%
               {\includegraphics{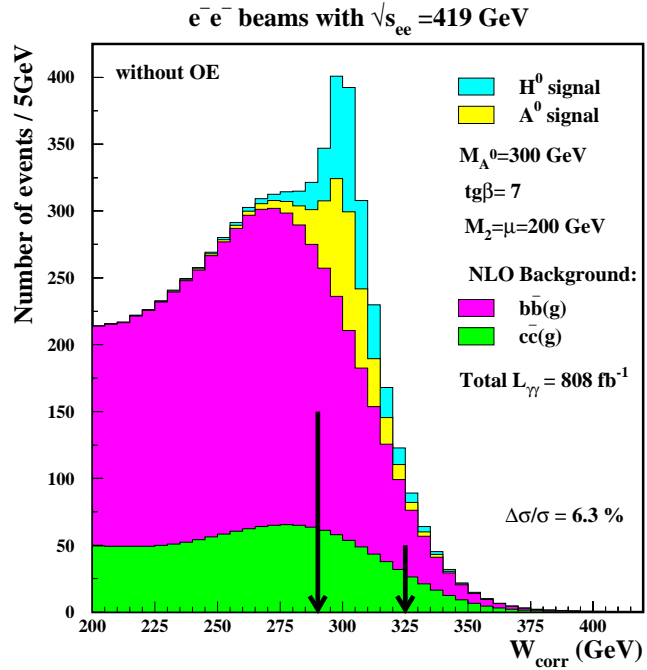}} \par}
{\centering \resizebox*{!}{\figheightDm}%
               {\includegraphics{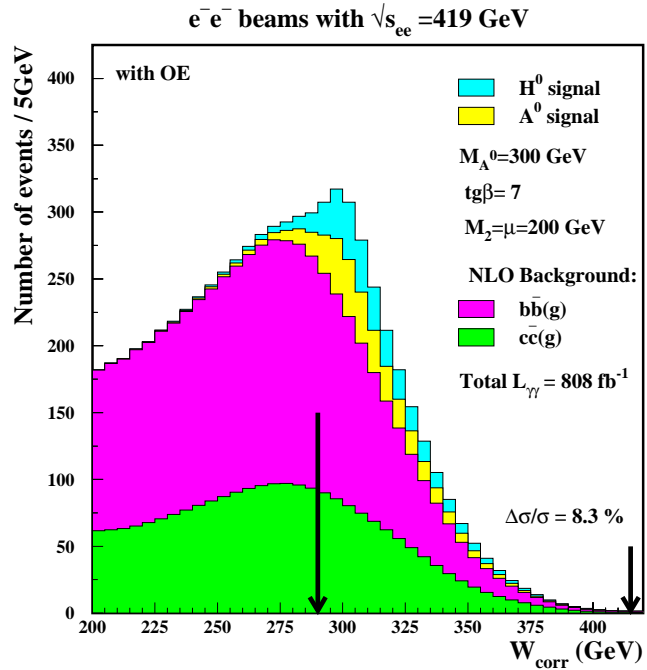}} \par}

\caption{\label{fig:Wcorr}
As in Figs.\ \ref{fig:ResultWithNLOBackgd}, for the corrected invariant-mass, \protect\( W_{corr}\protect \),
distributions. \newline
%
%{\bf Upper plot}: overlaying events (OE) are not included;
%arrows indicate the mass window optimized for the measurement of the 
%$\sgagaAHbb$, leading to the statistical precision of~6.3\%. 
% 
%{\bf Lower plot}: overlaying events (OE) are included;
%the statistical precision is~8.3\%.
The statistical precision is 6.3\% without overlaying events (upper plot)  
and 8.3\% with overlaying events (lower plot).
}
\end{figure}

The  \( W_{corr} \) distributions, obtained for the  signal and
background events,  are shown in Figs.\ \ref{fig:Wcorr}. 
Results obtained before (upper plot) and after (lower plot)
taking into account overlaying events (OE) are compared.
If overlaying events are included, 
the most precise measurement of the Higgs-boson  cross section
is obtained for  the   \( W_{corr} \) mass window
between 290 and 415 GeV, as indicated by arrows.
In the selected \( W_{corr} \) region one expects, after one year of
the Photon Collider running at the  nominal luminosity,
about 610 reconstructed signal
events and 2000  background events  (\ie \( S/B \approx 0.3 \)).
This corresponds to the measurement with the expected relative 
statistical precision of 8.3\%.

%\pagebreak

\begin{figure}[thb]
% \vspace{-1cm}
{\centering \resizebox*{!}{\figheight}%
               {\includegraphics{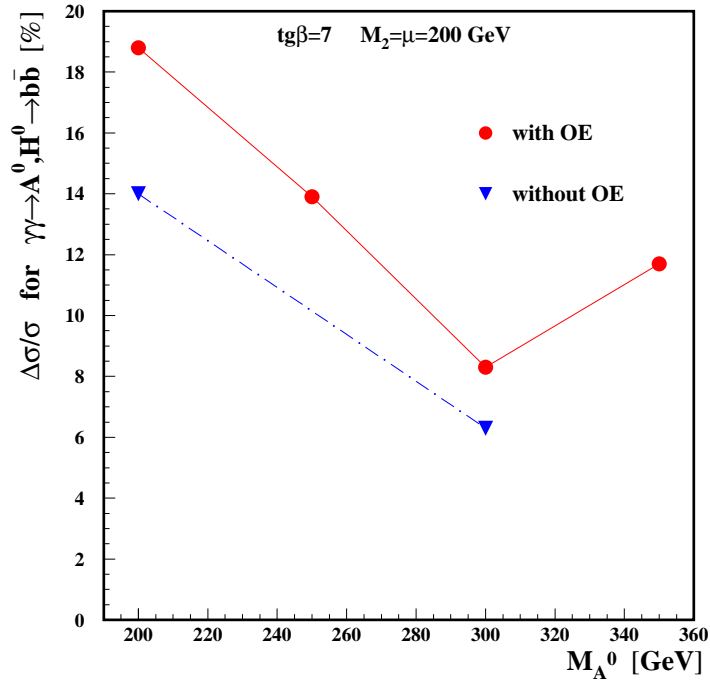}} \par}

\caption{\label{fig:PrecisionSummary}
Precisions of $\sgagaAHbb$ measurement are shown  for \MAOeq 200-350 GeV and \MSSMpars; 
with and without overlaying events (OE), as indicated in the plot.
The points are connected with lines to guide the eye. 
}
\end{figure}

We have performed also a full simulation of the signal and background events 
for \MAOeq 200, 250 and 350 GeV, choosing for each mass 
an optimal $\emem$ beam energy.
%Precision of $\sgagaAHbb$ measurement in each case was calculated.
In Fig.\ \ref{fig:PrecisionSummary} a statistical precision of $\sgagaAHbb$ measurement 
for all considered masses are presented. 
For comparison also our earlier estimates \cite{NZKamsterdam}, 
obtained without overlaying events, are shown.

\section{Conclusions}

In this paper we present results of a full simulation of a signal 
due to the heavy MSSM Higgs bosons $\AO$ and $\HO$ decaying into $\bbbar$  
and background events 
for the Photon Collider at TESLA. 
We study masses \MAOeq 200, 250, 300 and 350 GeV,
and for each mass we choose an optimal $\emem$ beam energy. 
Following \cite{MMuhlleitner}, we study 
parameters of the MSSM, which correspond to the ``LHC wedge'',
for which $\AO$ and $\HO$ are almost   degenerate in mass.
%
% which differ by $\sim 1$ GeV and they dominantly decay into  $\bbbar$. 
%We apply various cuts, and take into account an effect of overlaying events as in a light  SM Higgs-boson analysis.
%
We performed a realistic simulation 
with the NLO background, corrections for escaping neutrinos, 
with realistic \btagging{} and taking into account overlaying events,
as in our SM-Higgs analysis \cite{NZKSMeps2003}.

Our analysis shows that, for the MSSM Higgs-bosons at $\MAO \sim$ 300 GeV,
the cross section $\sgagaAHbb$ can be measured
with a statistical precision around 8\%, 
for other masses it is lower -- from 10\% till 20\%.
Although this result is less optimistic than the earlier estimate \cite{MMuhlleitner},
it is  still true that a photon--photon collider gives opportunity
of a precision measurement of $\sgagaAHbb$, 
assuming that we know the mass of a Higgs boson(s). 
The issue arises how  to distinguish (separate) $\AO$ from $\HO$.
To get a reliable answer to this question an additional study is needed.
We  confirm results of \cite{JikiaAndSoldner} which indicated that 
the reconstructed mass resolutions (without including overlaying events)
are greater than $\sim 2$ GeV.
A discovery of  MSSM Higgs-bosons requires energy scanning or a run with 
a broad luminosity spectrum, perhaps  followed by the run with a peaked one 
\cite{Asner}.
%
% as assumed in \cite{MMuhlleitner}.

We conclude, that there is still a room for an optimization of $\theta_{min}$ cut
to  minimize an influence of overlaying events, 
which may increase a statistical precision.

\newpage

%%%
%%%
% 
%
%%%
%%%

\newpage

\newpage

%%%
%%%
\newpage

%%%
%%%

\subsection*{Acknowledgments}
We would like to thank M.~M.~M\"{u}hlleitner, M.~Spira and P.~Zerwas for valuable discussions.
M.K.~acknowledges partial
support by the Polish Committee for Scientific Research, Grants 2 P03B 05119 (2003), 
5 P03B 12120 (2003), and by the European Community's
Human Potential Programme under contract HPRN-CT-2000-00149 Physics
at Colliders.

\end{document}